\documentclass{emulateapj}
\usepackage{amssymb}
\usepackage{amsmath}

\newcommand{\loiii}{\ensuremath{L_{\rm{[OIII]}}}\,}
\newcommand{\ewoiii}{\ensuremath{\rm{EW}_{\rm{[OIII]}}}\,}

\newcommand{\mbh}{\ensuremath{M_{\rm{BH}}}\,}

\received{July 23, 2017}
\revised{September 6, 2017}
\accepted{October 3, 2017}

\shorttitle{Evidence for higher black hole spin in radio-loud quasars}
\shortauthors{Schulze et al.}

\begin{document}

\title{Evidence for higher black hole spin in radio-loud quasars}

\author{Andreas Schulze\altaffilmark{1,7}, Chris Done\altaffilmark{2}, Youjun Lu\altaffilmark{3,4}, Fupeng Zhang\altaffilmark{5}, Yoshiyuki Inoue\altaffilmark{6}}
\email{E-mail: andreas.schulze@nao.ac.jp}

\altaffiltext{1}{National Astronomical Observatory of Japan, Mitaka, Tokyo 181-8588, Japan}
\altaffiltext{2}{Centre for Extragalactic Astronomy, Department of Physics, University of Durham, South Road, Durham, DH1 3LE, UK}
\altaffiltext{3}{National Astronomical Observatories, Chinese Academy of Sciences, Beijing 100012, Chin}
\altaffiltext{4}{School of Astronomy and Space Sciences, University of Chinese Academy of Sciences, No. 19A Yuquan Road, Beijing 100049, China}
\altaffiltext{5}{School of Physics and Astronomy, Sun Yat-Sen University, Guangzhou 510275, China}
\altaffiltext{6}{Institute of Space and Astronautical Science JAXA, 3-1-1 Yoshinodai, Chuo-ku, Sagamihara, Kanagawa 252-5210, Japan}
\altaffiltext{7}{EACOA Fellow}



\begin{abstract}
One of the major unsolved questions on the understanding of the AGN population is the origin of the dichotomy between radio-quiet and radio-loud quasars. The most promising explanation is provided by the spin paradigm, which suggests radio-loud quasars have higher black hole spin. However, the measurement of black hole spin remains extremely challenging.
We here aim at comparing the mean radiative efficiencies of carefully matched samples of radio-loud and radio-quiet SDSS quasars at $0.3<z<0.8$. 
We use the [\ion{O}{3}] luminosity as an indirect average tracer of the ionizing continuum in the extreme-UV regime where differences in the SED due to black hole spin are most pronounced. We find that the radio-loud sample shows an enhancement in [\ion{O}{3}] line strength by a factor of at least 1.5 compared to a radio-quiet sample matched in redshift, black hole mass and optical continuum luminosity or accretion rate. We argue that this enhancement is caused by differences in the SED, suggesting higher average bolometric luminosities at fixed accretion rate in the radio-loud population. This suggests that the radio-loud quasar population has on average systematically larger radiative efficiencies and therefore higher black hole spin than the radio-quiet population, providing observational support for the black hole spin paradigm.
\end{abstract}

\keywords{Galaxies: active - Galaxies: nuclei - quasars: general}



\section{Introduction}

Quasars constitute the most luminous types of Active Galactic Nuclei (AGN), where a supermassive black hole (SMBH) is powered by significant mass accretion through a thin accretion disc. The first quasars where initially discovered via their radio emission \citep{Matthews:1963,Schmidt:1963} a fact still reflected in their name (quasar="quasi stellar radio source"). However, only about 10\% of all quasars are radio-loud (RL), i.e. have relativistic jets with high bulk Lorentz factor, $\Gamma\sim 10$. The majority have much weaker core radio emission, so are termed radio--quiet (RQ). This distinction in radio to optical flux is not sharp, but there are clearly two populations  \citep[e.g.][]{Kellermann:1989,Ivezic:2002,Balokovic:2012}.
Viewing angle with respect to the jet will change the observed intensity due to relativistic beaming \citep{Scheuer:1979}, 
but it is now clear that inclination unifies different classes of RL AGN rather than explaining the difference between RL and RQ \citep[e.g.][]{Urry:1995}.

The physical origin of this distinction in jet behaviour is one of the main unsolved problems in AGN physics. The accretion flow and its associated jet should be completely determined by the fundamental parameters of the SMBH, its mass and spin, and the mass accretion rate onto the black hole. There are  dependancies of the radio-loud fraction (RLF) on redshift and luminosity \citep{Jiang:2007}\footnote{However, as argued by \citet{Kratzer:2015}, the depth of the radio data is not sufficient to robustly exclude that these trends are due to incompleteness or some other selection effect.}, but these are probably more fundamentally associated with changes in black hole mass $M_{\rm{BH}}$ \citep{Laor:2000,Lacy:2001,Dunlop:2003,Chiaberge:2011,Kratzer:2015,Coziol:2017} and normalized accretion rate, i.e Eddington ratio $L_{\rm{bol}}/L_{\rm{Edd}}$ \citep{Lacy:2001,Ho:2002,Sikora:2007,Coziol:2017} over cosmic time.

The dependence of the jet on Eddington ratio can be seen explicitly in the stellar mass black hole binary systems (BHB). These show a distinct spectral transition as the luminosity drops below $L_{bol}/L_{Edd}\sim 0.02$ which is best explained by the accretion flow changing from a geometrically thin, optically thick, cool disc \citep{Shakura:1973} (disc dominated state) to a geometrically thick, optically thin, hot flow \citep{Narayan:1995,Esin:1997,Done:2007} (Compton dominated state). A compact flat spectrum jet is seen in the hot flow states, with radio emission which is (generally) proportional to $L^{0.7}$ when the source is in the Compton dominated state, but the jet emission collapses as the source makes the transition to the disc dominated state \citep{Fender:2004, Corbel:2013}. This motivated the suggestion that the RL and RQ AGN can be explained by the same transition in accretion flow properties \citep{Maccarone:2003,Sikora:2007}, but the jet behaviour in the BHB is never really analogous to that of RL AGN as the BHB radio jets cannot be highly relativisitic. This would introduce too much scatter into the radio-X-ray relation across the different inclination angles of BHB
\citep{Heinz:2004,Done:2016}.

Thus $L_{\rm{bol}}/L_{\rm{Edd}}$  alone is not sufficient to explain the radio dichotomy, and jet launching is a plasma process so should be scale invariant with respect to $M_{\rm{BH}}$ (though its radio emission at a single frequency should scale with mass due to self absorption \citep{Merloni:2003}). Instead, a more plausible scenario is the so-called "spin paradigm" \citep{Blandford:1990,Wilson:1995}. Theoretical models show that a relativistic radio jet can be generated from the rotational energy of a spinning black hole via magnetic fields brought into the ergosphere by the accretion flow \citep{Blandford:1977}.  This scenario associated RL quasars with rapidly spinning SMBHs, while RQ quasars should have low spin \citep[e.g.][]{Maraschi:2012}. 

However, this simple picture was discarded after detailed X-ray spectroscopic studies of nearby AGN resulted in high spin for many RQ objects \citep[e.g. the review by][and references therein]{Reynolds:2014}. While these studies also find low/intermediate spin values for several AGN, in particular for SMBH masses around $10^8\, M_\odot$ \citep[e.g.][]{Patrick:2011,Walton:2013}, the high spin values reported for some RQ AGN argues against BH spin as the sole factor responsible for producing a RL AGN. Instead, newer models have concentrated on non-deterministic factors such as the long term history of accretion of net magnetic flux onto the black hole \citep{Sikora:2013}, which could result in a magnetically arrested disc producing a powerful jet \citep[MAD:][]{Narayan:2003,McKinney:2012,Sadowski:2017}.  
Nonetheless, there are still some significant uncertainties in the spin measurements from X-ray spectroscopy.  In particular, many of the highest spin RQ objects (i.e. the ones which most challenge the spin-jet paradigm) 
are low mass, high mass accretion rate AGN (Narrow Line Seyfert 1s) with  super Eddington accretion flows. This
means that the key assumption of the iron line modeling of a clean line of sight view to a flat disk may not hold \citep{Done:2016}.

In view of this uncertainty, it is important to have an alternative method to estimate SMBH spin. This is provided by the radiative efficiency $\epsilon$ of the AGN, which connects the bolometric accretion disc luminosity output to the mass accretion rate,
\begin{equation}
L_{\rm{bol}}= \epsilon \dot{M}_{\rm{acc}} c^2 \ .
\label{eq:lbol}
\end{equation}
This radiative efficiency is set by the innermost stable circular orbit (ISCO) of the black hole, beyond which any material falls into the black hole without losing further energy. This radius depends solely on the black hole spin \citep{Novikov:1973}, parametrized by a dimensionless parameter of $a$ (where $-1\leq a <1$, with negative values indicating a retrograde disc). Based on $a$, the radiative efficiency can vary between $\epsilon = 0$ for maximally retrograde spin, to $\epsilon=0.054$ for a non-spinning black hole and $\epsilon=0.42$ for a maximally spinning Kerr black hole \citep{Shapiro:1983}.

An average value for the radiative efficiency of the entire AGN population can be derived from continuity equation arguments based on evolving the AGN luminosity function and the black hole mass function \citep{Soltan:1982,Yu:2002,Yu:2008,Shankar:2009}.  These studies suggest $\epsilon\sim0.1$ on average \citep{Marconi:2004,Zhang:2012,Ueda:2014}, corresponding to a moderate spin of $a=0.67$.

The accretion rate can be determined with reasonable accuracy from the observed optical flux from the outer thin disc emission once the black hole mass is known (modulo inclination angle) \citep[e.g.][]{Davis:2011}. The true bolometric luminosity is difficult to assess for individual sources, since the spectral energy distribution (SED) usually peaks in the far- to extreme-UV (EUV) regime which cannot be directly observed due to absorption in our Galaxy. 
While for individual objects this leads to significant uncertainties \citep[e.g.][]{Raimundo:2012}, insight into the AGN population as a whole can be gained nevertheless either by adopting SED templates based on empirical relations between spectral slopes and luminosities obtained from multi-wavelength observations for all sources \citep{Davis:2011,Wu:2013} or by bracketing the uncertainty in the bolometric corrections \citep{Netzer:2014,Trakhtenbrot:2014}. These studies for example confirm an average value for the radiative efficiency of 0.1 \citep{Davis:2011,Wu:2013}.
Given sufficient coverage of the accretion disc SED it is also possible to constrain the spin directly from disc continuum fitting for certain objects \citep[e.g.][]{Done:2013,Capellupo:2016,Capellupo:2017}.

\begin{figure}
\centering
\resizebox{\hsize}{!}{\includegraphics[clip]{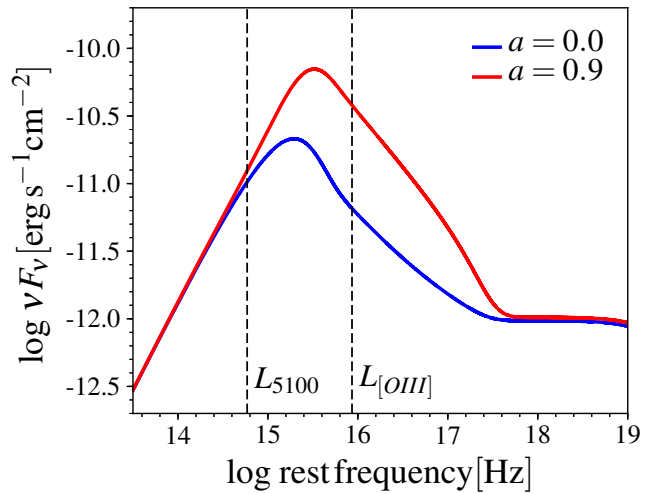}}
\caption{Illustration of the typical spectral energy distribution of AGN derived from accretion disc models (including soft X-ray excess and hard X-ray corona) for two extreme cases of black hole spin: zero spin $a=0$ (blue line) and high spin $a=0.9$ (red line). For both cases the rest-frame optical and X-ray emission can be the same and the spin difference only manifests itself in the EUV regime.}
\label{fig:sed}
\end{figure}

Constraining the accretion power at fixed rest frame optical luminosity serves as a direct probe of SMBH spin. 
However, the SED is significantly more complex than expected from simple accretion disc models. Instead, it can be fit fairly well by assuming that there is a radial transition between the outer standard, blackbody emission disc to an inner region where the energy does not quite thermalize, so is emitted instead as low temperature, optically thick Comptonisation (soft X-ray excess) in addition to the high temperature, optically thin Comptonisation in a corona \citep{Done:2012}.  The effect of black hole spin on this composite structure is not possible to completely specify as their underlying physics is not well understood. However, the additional accretion energy associated with increasing black hole spin derives from the decreasing radius of the last stable circular orbit around the black hole. Thus the additional power is released at the smallest radii, so should have no impact on the outer standard disc emission, but instead should be concentrated in the soft X-ray excess and coronal components. Fig.~\ref{fig:sed} shows a potential SED which could result from a black hole of mass $10^9 M_\odot$ with $\dot{M}=8.5\, M_\odot/$yr for spin a=0 (blue) and 0.9 (red). Here we have assumed that all of the additional accretion power is released in the (mostly unobservable) soft X-ray excess component i.e. the maximum change in ionizing luminosity for the minimum change in observed flux. In this illustrative case the difference in spin is only reflected in enhanced emission in the EUV.

We here propose to use an indirect tracer of the EUV luminosity, namely the [\ion{O}{3}] luminosity to address the question if RL and RQ quasars have different average radiative efficiencies and therefore different mean black hole spin.
The [\ion{O}{3}] $\lambda$5007\AA{} emission line has an ionization potential of 35.6eV (354\AA), i.e. is ionized by radiation close to the SED peak. For realistic AGN SEDs, the ionizing luminosity traces the bolometric luminosity approximately linearly. It therefore serves as a bolometric luminosity indicator, sensitive to black hole spin differences. The [\ion{O}{3}] line luminosity,  \loiii, is determined by the ionizing radiation field and the conditions in the Narrow Line Region (NLR) from where it is emitted (e.g. density, covering factor). Since there can be significant variation in the latter,  \loiii  may not be a precise predictor of the bolometric luminosity for individual objects \citep[e.g.][]{Baskin:2005,Stern:2012}, but can be used  as a tracer for the average population \citep[e.g.][]{Heckman:2004,Netzer:2009}.  We here make the assumption that RL and RQ quasars do not have systematically different physical conditions of their NLR. While this assumption is plausible, we note that it is currently observationally not well established. We will discuss potential ramifications of this assumption on our results further below.
In this case, RL and RQ quasar samples matched in optical continuum luminosity and black hole mass should have consistent mean \loiii if they have on average the same radiative efficiency, thus black hole spin. This is the hypothesis we aim to test in this paper.

\section{Sample}
We use the Sloan Digital Sky Survey (SDSS) DR7 quasar catalog \citep{Schneider:2010,Shen:2011}, consisting of unobscured, luminous AGN (having at least one broad emission line $>1000$~km s$^{-1}$ and $i$-band magnitude $M_i<-22$). Of these we focus on the subset targeted by a homogeneous optical color selection method \citep{Richards:2002} and restrict the redshift range to $0.3<z<0.84$, to have the H$\beta$ and [\ion{O}{3}] emission lines covered by the SDSS spectra. Their radio properties are taken from the Faint Images of the Radio Sky at Twenty Centimeters \citep[FIRST;][]{Becker:1995} survey. We thus restrict the SDSS sample to the FIRST footprint. The FIRST identifications and radio properties for the quasars are taken from the SDSS DR7 catalog from \citet{Shen:2011}, based on a matching algorithm outlined in \citet{Jiang:2007}, using an initial matching radius of 30\arcsec, which is reduced to 5\arcsec\ if only one FIRST source is detected within this radius. We define a radio-loud quasar via the radio loudness parameter $R=f_{6\rm{cm}}/f_{2500}$, the rest-frame ratio of the radio flux density at $6$~cm (5~GHz) and the optical flux density at 2500~\AA{}. The rest-frame flux density at 6~cm is determined from the observed flux density at 20~cm, assuming a power law with spectral slope $\alpha_\nu=-0.5$. The optical flux density $f_{2500}$ is obtained from the optical spectrum. As our default definition of a RL quasar we adopt $R>R_{\rm{crit}}$ with $R_{\rm{crit}}=10$ \citep{Kellermann:1989}, but will also explore other definitions of radio-loudness to evaluate the robustness of our results against this specific choice.

The FIRST survey has a radio flux density limit of about 1.0~mJy \citep{Becker:1995}. For this radio flux limit the criteria $R>10$ corresponds to an optical magnitude limit of $i<18.9$~mag \citep[see][]{Jiang:2007,Kratzer:2015}, slightly brighter than the magnitude limit for our SDSS parent sample at  $i=19.1$~mag. Above this optical flux limit FIRST is complete to AGN with $R>10$, so we can uniquely classify each quasar into RL or RQ.
These selection criteria result in an initial sample of 8054 quasars.

\begin{figure*}[tbh]
\centering
\includegraphics[width=7cm,clip]{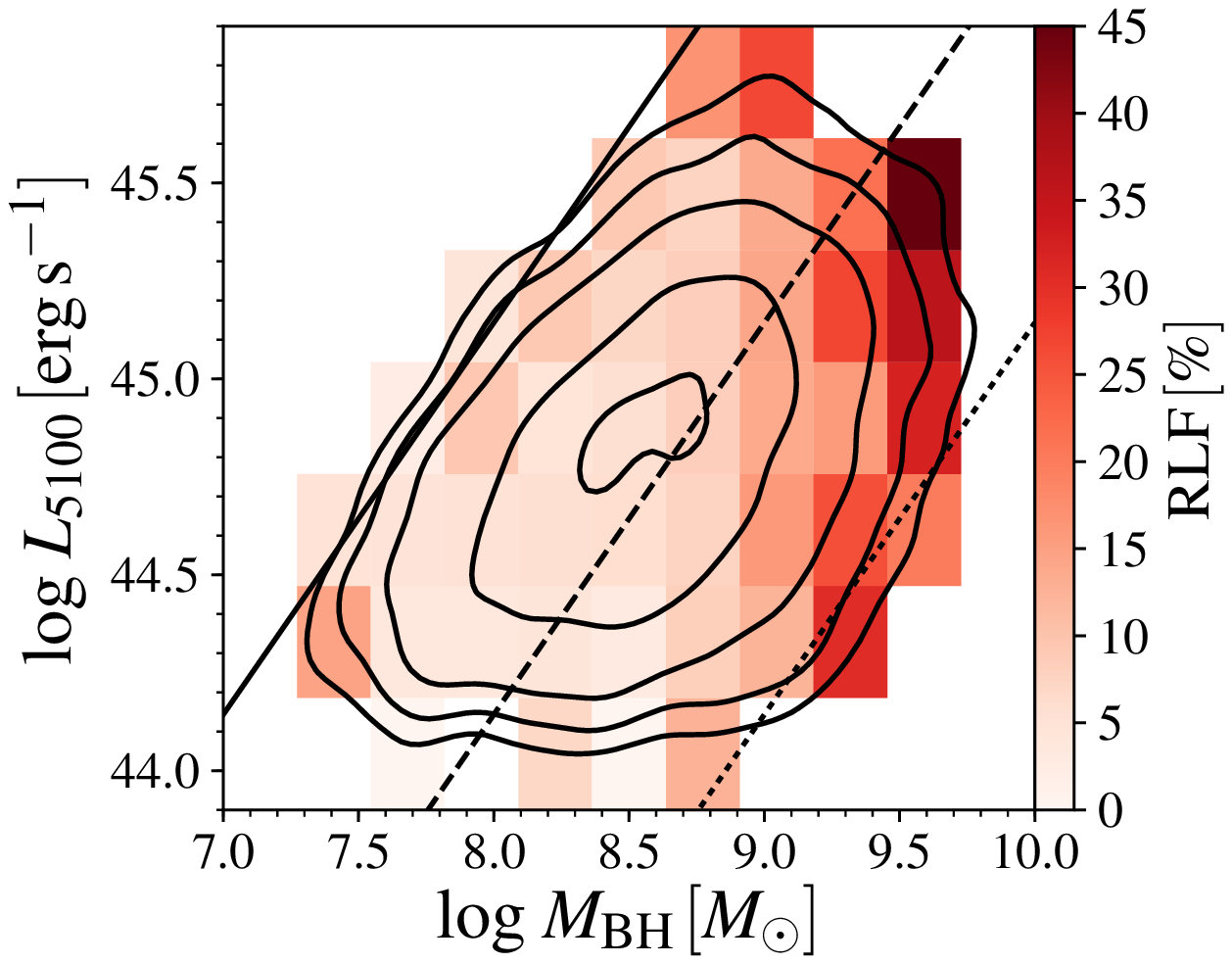} \hspace{0.5cm}
\includegraphics[width=7cm,clip]{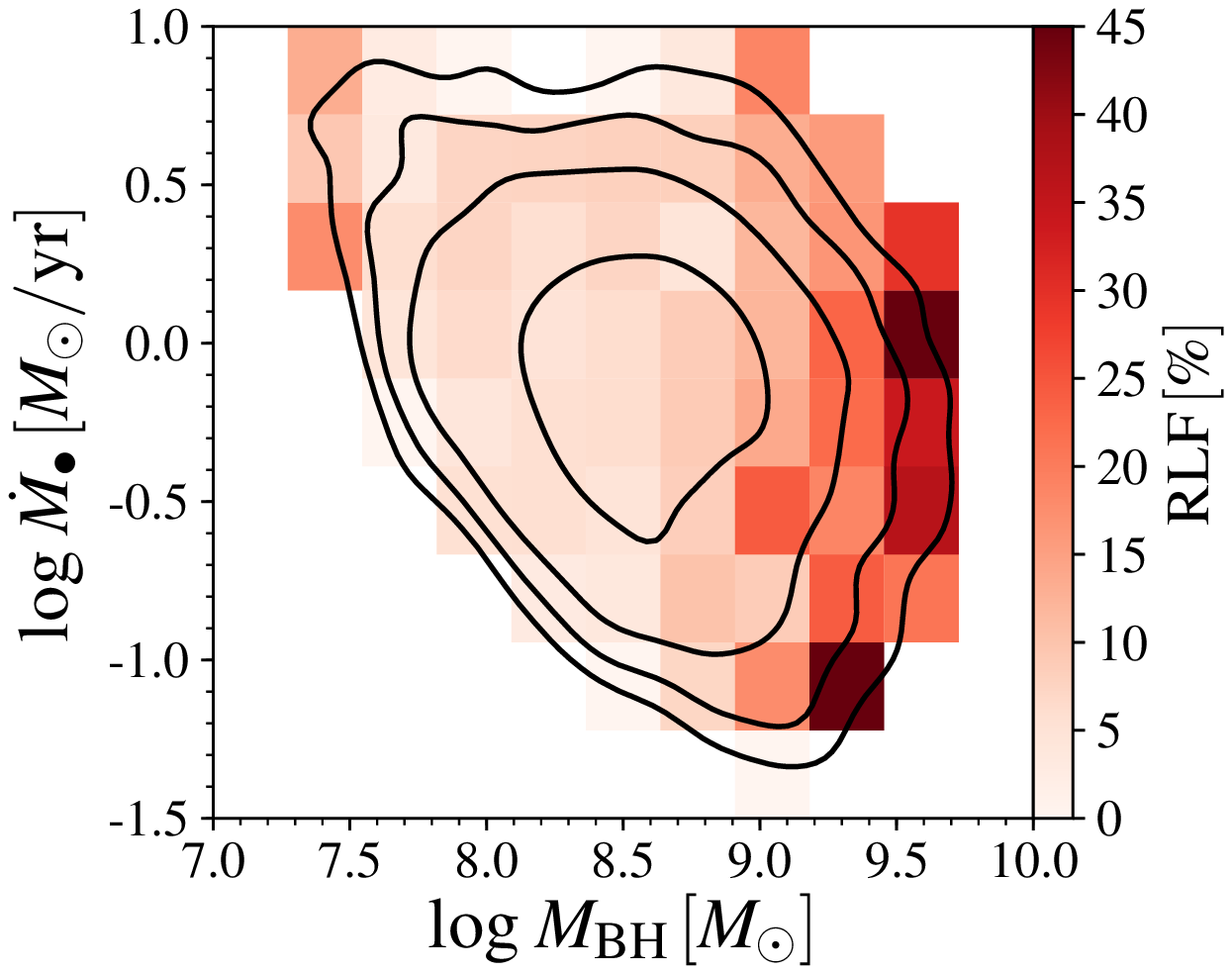}
\caption{Radio-loud fraction in the SDSS $0.3<z<0.84$ quasar sample as a function of black hole mass and continuum luminosity at 5100\AA{} (left panel) and black hole mass and accretion rate (right panel). The contours show the number density of the full sample within the parameter space. The solid, dashed and dotted lines in the left panel indicate Eddington ratios of 100, 10 and 1\%, assuming a bolometric correction of $L_{ bol}=9.26 L_{5100}$ \citep{Shen:2008}.}
\label{fig:rlf}
\end{figure*}

We utilize the quasar emission line and continuum measurements in the catalog from \citet{Shen:2011} and the black hole mass estimates derived from these using the virial method \citep[e.g.][]{McLure:2004,Vestergaard:2006}. We only use black hole mass estimates obtained from the broad H$\beta$ line, which is considered to provide the most robust mass estimator, since it is directly calibrated to reverberation mapping \citep[e.g.][]{Peterson:2004,Vestergaard:2006,Bentz:2009}. Specifically, we use full width half-maximum (FWHM) and continuum luminosities $L_{5100}$ measured by \citet{Shen:2011} and the virial relation by \citet{Vestergaard:2006}:
\begin{equation}
\mbh (\rm{H}\beta)= 10^{6.91} \left( \frac{L_{5100}}{10^{44}\,\mathrm{erg\,s}^{-1}}\right)^{0.50} \left( \frac{\mathrm{FWHM}}{3000\,\mathrm{km\,s}^{-1} }\right)^2 M_\odot  \label{eq:mbhHb}
\end{equation} 

We exclude objects with a mass measurement error $>0.5$~dex, as derived in \citet{Shen:2011} from Monte Carlo simulations. This gives a final sample of 7788 quasars. Of these 746 are classified as RL and 7042 as RQ, approximately consistent with the standard $~10$\% RL quasar fraction. The vast majority of our RL sample has radio luminosities $L_{1.4\rm{GHz}}>10^{24}$~W~Hz$^{-1}$.

Mass accretion rates $\dot{M}_{\rm{acc}}$ for this sample have been derived in \citet{Wu:2013}. 
The rest-frame optical to UV radiation of quasars is emitted from an optically thick, geometrically thin accretion disc \citep[e.g.][]{Shakura:1973}, whose observed spectrum is mainly determined by the mass of the supermassive black hole, their spin, the mass accretion rate $\dot{M}_{\rm{acc}}$ and the inclination angle $i$ of the disc towards the line of sight \citep[e.g.][]{Novikov:1973,Krolik:1999}. The role of BH spin here is mainly to set the inner radius of the accretion disc. However, the optical luminosity, e.g. at 5100~\AA{}, is dominated by emission from the outer disc and therefore is less affected by the BH spin \citep[e.g.][]{Davis:2011}, as illustrated in Fig.~\ref{fig:sed}. We can therefore use the measured optical continuum luminosity at 5100~\AA\ $L_{5100}$ and the BH mass $M_{\rm{BH}}$, derived from the virial method, to estimate the mass accretion rate with reasonable accuracy, without prior knowledge of the BH spin. To good accuracy the accretion rate derived from disc models scales like  $\dot{M}_{\rm{acc}}\propto L_{\rm{opt}}^{3/2}\mbh^{-1}$ \citep[e.g.][]{Collin:2002,Davis:2011}.
In detail, \citet{Wu:2013} used the thin accretion disc model TLUSTY \citep{Hubeny:2000}, fixing the BH spin to $a=0.67$ (corresponding to $\epsilon\sim0.1$) and the inclination of the disc to $\cos i=0.8$ to obtain an estimate of $\dot{M}_{\rm{acc}}$ for every quasar. These values suffer from uncertainties due to the use of virial BH masses instead of their true masses and by fixing the inclination angle. However, to first order these possible biases should be the same for both the RL and RQ sample. We emphazise that the derived $\dot{M}_{\rm{acc}}$ is basically independent of the specific choice of spin $a$ in the models, since we trace the luminosity through the outer disc, so the SMBH spin assumed here does not have an effect on our discussions below.

\section{Results}
\subsection{RLF dependence on black hole mass, luminosity and accretion rate}
We briefly review the dependencies of the radio-loud fraction (RLF) on black hole mass, luminosity and accretion rate for our sample. The results are shown in Fig.~\ref{fig:rlf}, where we compute the fraction of radio-loud QSOs in bins of $M_{\rm{BH}}$ and $L_{5100}$ and of $M_{\rm{BH}}$  and $\dot{M}_{\rm{acc}}$, respectively. We do not take into account the detailed completeness of FIRST close to the detection limit \citep{Kratzer:2015} and may also miss a small number of very extended sources \citep{Lu:2007}. Thus Fig.~\ref{fig:rlf} should rather highlight the main trends for our sample.

We find that the RLF for unobscured quasars shows a clear dependence on $M_{\rm{BH}}$, with the most massive black holes having the highest fraction of RL quasars, consistent with previous work \citep{Laor:2000,Lacy:2001,Dunlop:2003,Kratzer:2015}. Both optical continuum luminosity $L_{5100}$  and accretion rate show a weaker trend. However, even at the extreme end of the RLF, for the most massive black holes, the majority of quasars are still radio-quiet. This indicates that a high $M_{\rm{BH}}$ likely supports the presence of a radio-jet, but is not sufficient to produce a RL quasar. At a given $M_{\rm{BH}}$ and $\dot{M}_{\rm{acc}}$ there must be an additional parameter at work, likely SMBH spin.

\begin{figure}
\centering
\resizebox{\hsize}{!}{\includegraphics[clip]{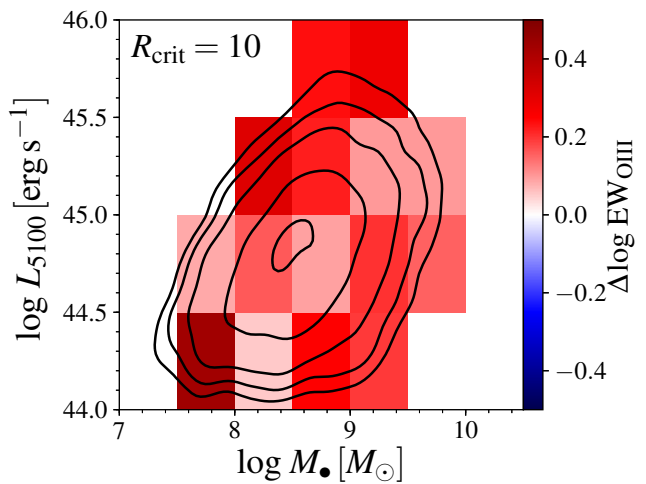}}
\caption{Difference in [\ion{O}{3}]  equivalent width, $\Delta \log \ewoiii=\log\ewoiii\rm{(RL)}-\log\ewoiii\rm{(RQ)}$ between the RL and RQ quasar sample in bins of constant black hole mass and optical continuum luminosity at 5100\AA{} $L_{5100}$ for a radio-loudness definition of $R>10$. The contours show the number density for the full sample. We show bins with at least 10 objects per bin.}
\label{fig:mlplane}
\end{figure}

\begin{figure*}
\centering
\includegraphics[width=5.9cm,clip]{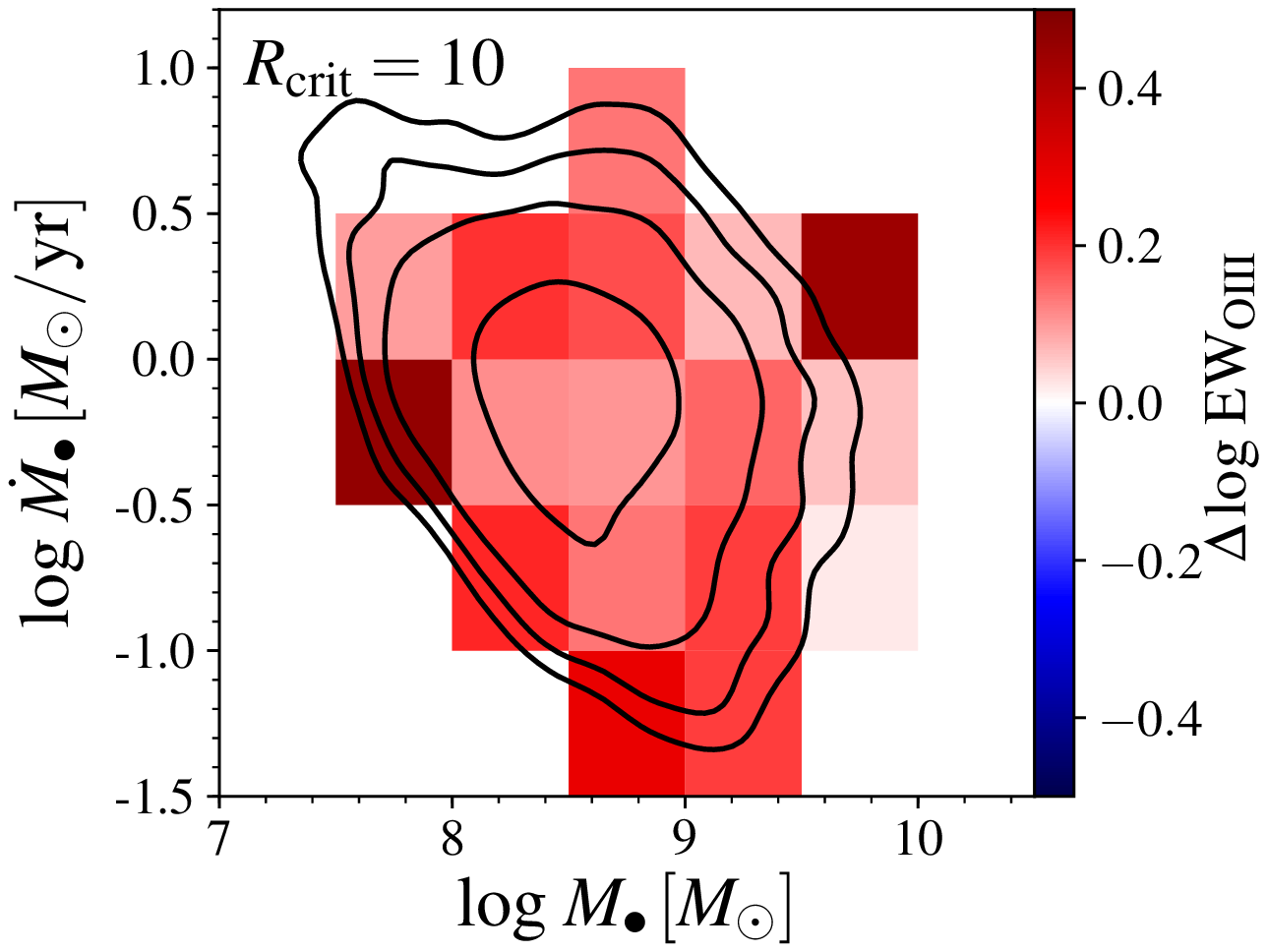}
\includegraphics[width=5.9cm,clip]{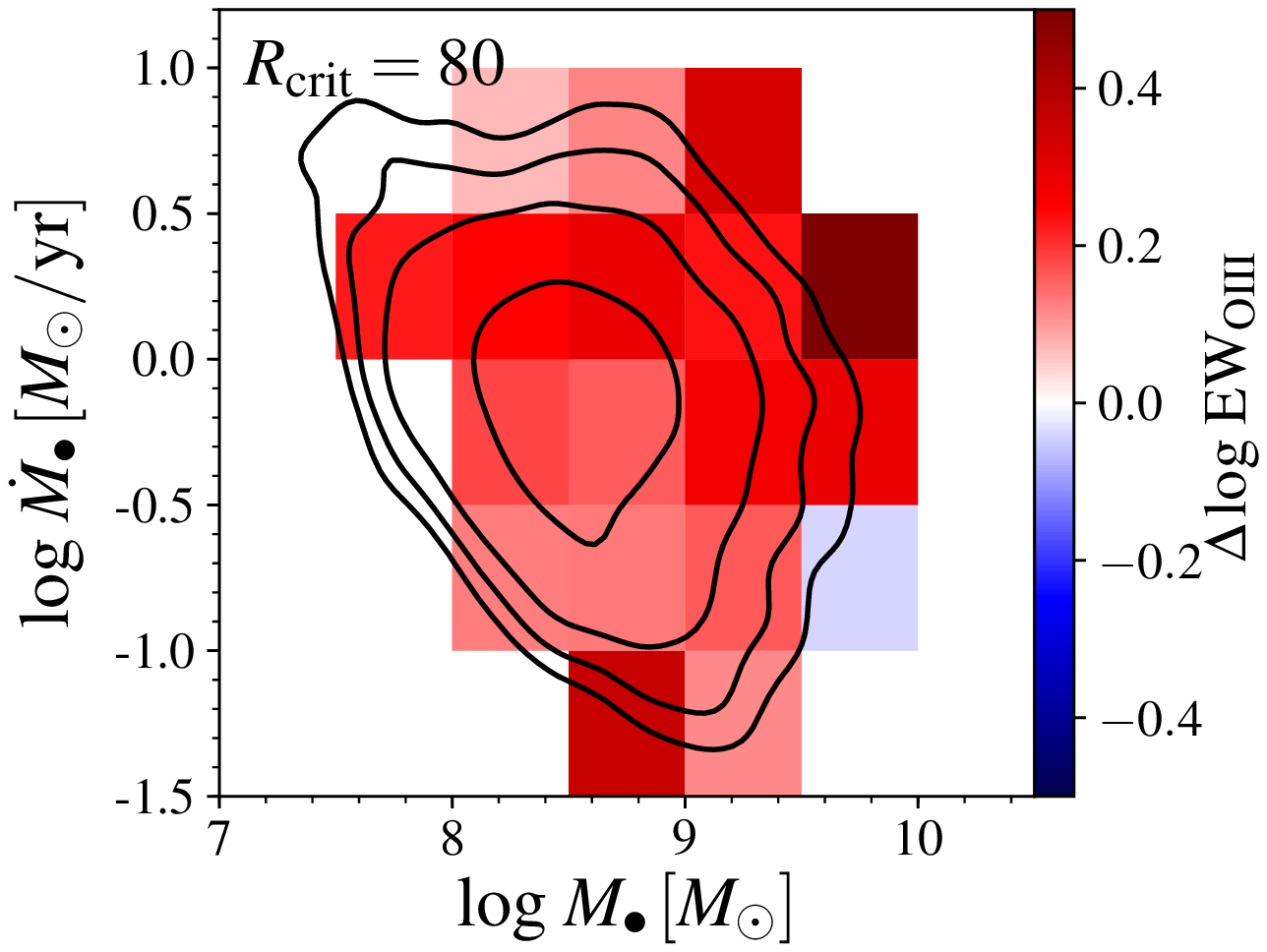}
\includegraphics[width=5.9cm,clip]{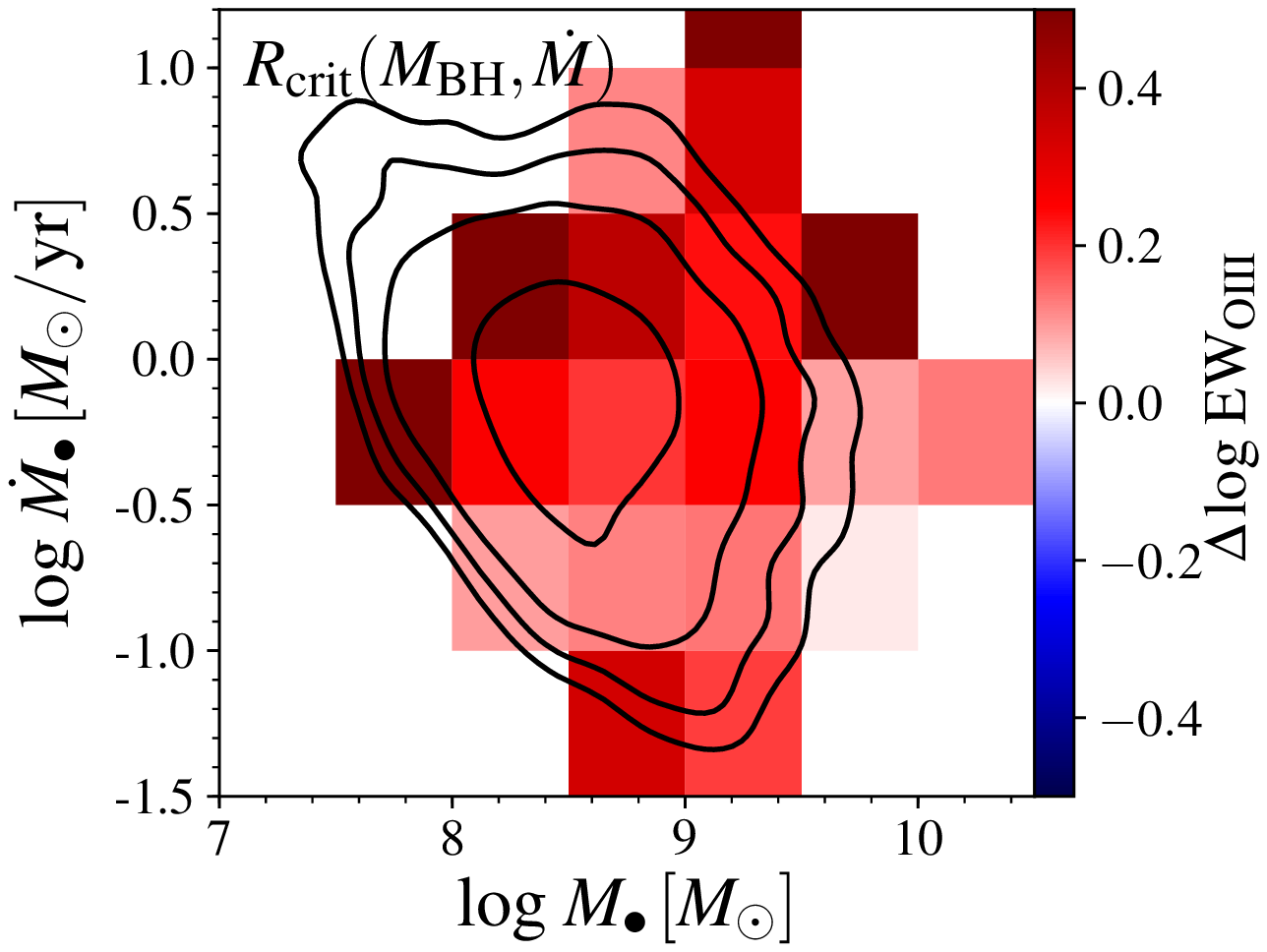}
\caption{Difference in [\ion{O}{3}]  equivalent width between the RL and RQ sample in bins of constant black hole mass and accretion rate $\dot{M}_{\rm{acc}}$. The contours show the number density for the full sample. We use definitions of radio-loudess of $R>10$ (left panel), $R>80$ (middle panel) and $R>R_{\rm{crit}}$ where $R_{\rm{crit}}$ depends on $M_{\rm{BH}}$ and $\dot{M}_{\rm{acc}}$ according to Eq.~\ref{eq:rfp}, i.e.   $R_{\rm{crit}}(M_{\rm{BH}},\dot{M})$ (right panel). We show bins with at least 10 (left), 4 (middle) and 2 (right panel) objects per bin.}
\label{fig:mmdotplane}
\end{figure*}

\begin{figure*}
\centering
\includegraphics[width=16cm,clip]{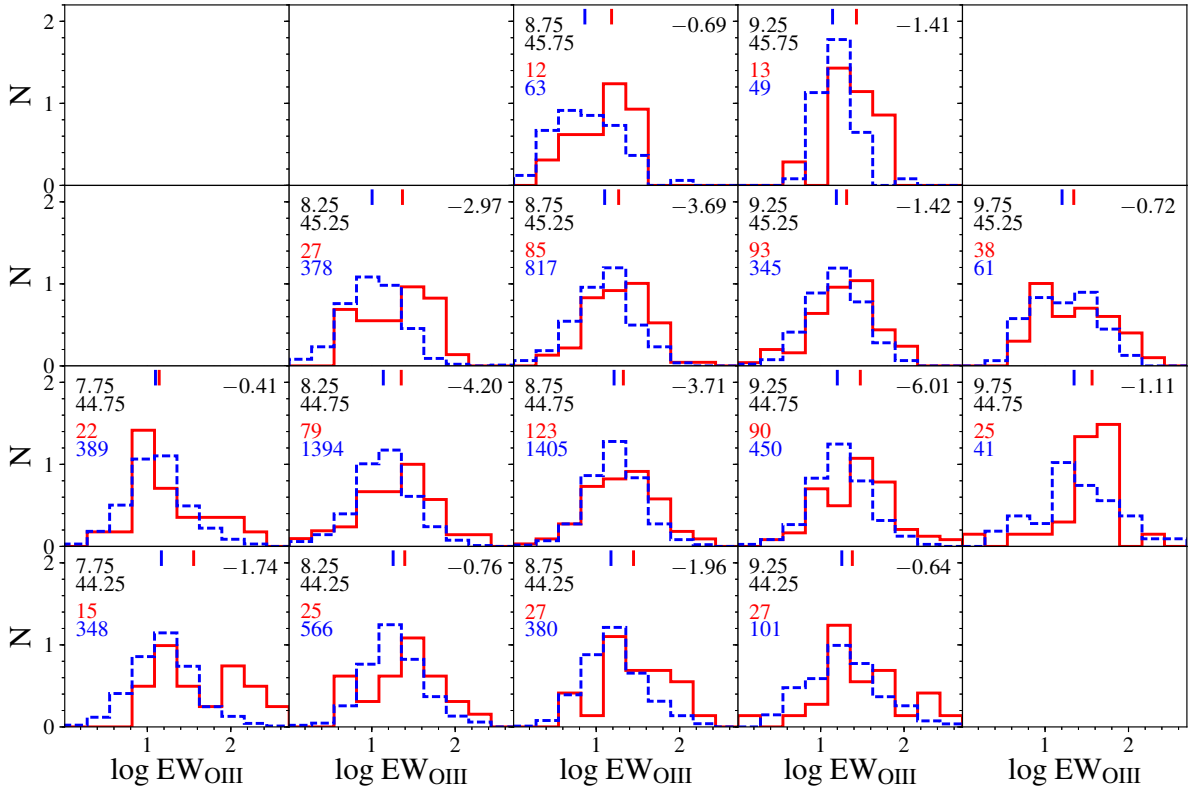}
\caption{Histogram of [\ion{O}{3}]  equivalent width for the RL (solid red lines) and the RQ (dashed blue lines) quasar sample in bins of $M_{\rm{BH}}$ and $L_{5100}$ (as listed in the upper left corners), matching the bins in Fig.~\ref{fig:mlplane}. The red and blue vertical markers indicate the mean equivalent width for the RL and RQ sample respectively. The numbers of RL (red) and RQ (blue) QSOs per bin are given on the middle right side. We also give the logarithm of the probability from a Kolmogorov-Smirnov test that the samples are drawn from the same distribution in the upper left corners.}
\label{fig:hist_bins}
\end{figure*}

\begin{figure*}
\centering
\includegraphics[width=16cm,clip]{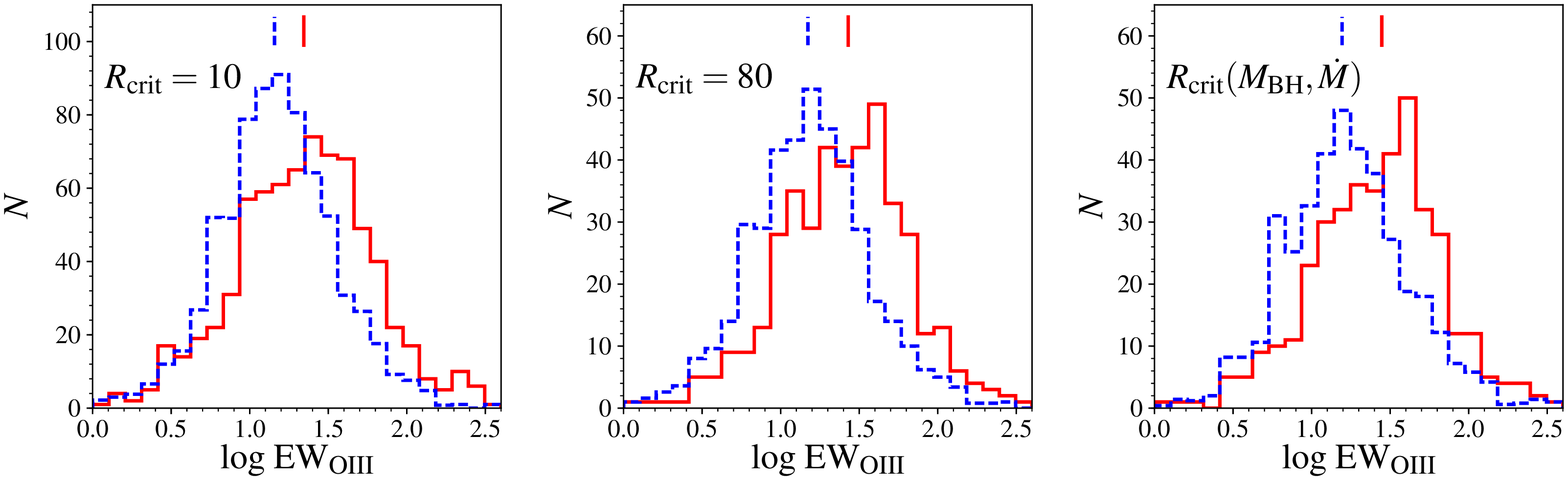}
\caption{Histogram of [\ion{O}{3}]  equivalent width for the RL quasar sample (solid red lines) and the matched RQ sample (dashed blue lines).The red and blue vertical markers indicate the mean equivalent width for the RL and RQ sample respectively. The source number on the y-axis is decreased for the RQ sample by a factor of 5 to match the RL sample size for better visualization. We use the same three different definitions of radio-loudness as in Fig.~\ref{fig:mmdotplane}.}
\label{fig:hist}
\end{figure*}

\subsection{Comparison of [\ion{O}{3}] emission line strengths}
For our test it is essential to control the RL and RQ quasar samples for the other fundamental SMBH parameters of black hole mass and accretion rate. This controls for the dependence of SED shape on $M_{\rm{BH}}$ and $\dot{M}_{\rm{acc}}$ and also for the mass dependence of the RLF shown above when comparing samples of RL and RQ quasars. 
It thus allows us to disentangle the known dependencies on mass and accretion rate so that any remaining difference is likely due to black hole spin assuming that both RQ and RL quasars sample the same range of inclination angles. It further ensures to be unaffected by luminosity trends like the known Baldwin effect in the [\ion{O}{3}] line \citep{Stern:2013,Zhang:2013}.
 
In the following we  control for  $M_{\rm{BH}}$ and $\dot{M}_{\rm{acc}}$ by either inspecting results in mass and luminosity/accretion rate bins or by matching the RL and RQ sample in these properties, finding consistent results. The question now is: Do these well matched RL and RQ quasar populations have a consistent average radiative efficiency?

To test this we investigate the [\ion{O}{3}] equivalent width, $\ewoiii$, distributions between the RL and matched RQ samples.
In Fig.~\ref{fig:mlplane} we show the difference in [\ion{O}{3}] equivalent width, $\Delta \log\ewoiii=\log\ewoiii\rm{(RL)}-\log\ewoiii\rm{(RQ)}$,
between our RL and RQ sub-samples in bins of constant $M_{\rm{BH}}$ and $L_{5100}$. It is evident that in all bins the [\ion{O}{3}] strength in the RL population is enhanced compared to the RQ population. Since  $M_{\rm{BH}}$ and $L_{5100}$ are the main parameters to determine $\dot{M}_{\rm{acc}}$ from the accretion disc model, this corresponds to a match in $M_{\rm{BH}}$ and $\dot{M}_{\rm{acc}}$. In the left panel of Fig.~\ref{fig:mmdotplane} we explicitly show that the RL population has higher mean \ewoiii in all bins of $M_{\rm{BH}}$ and $\dot{M}_{\rm{acc}}$.

In Fig.~\ref{fig:hist_bins} we directly compare the distribution of \ewoiii between both populations in the same bins of  $M_{\rm{BH}}$ and $L_{5100}$  as shown in Fig.~\ref{fig:mlplane}. There is a wide spread in \ewoiii for both populations in every bin due to the NLR properties. However, in every bin RL and RQ quasars show different distributions, as indicated by their mean \ewoiii. The variation of the mean \ewoiii across the bins is consistent with their statistical uncertainty in each individual bin. 
In most bins with a sufficient number of objects ($\gtrsim30$) these distributions for RL and RQ are statistically significantly different, according to a Kolmogorov-Smirnov test (with the log-probability that they are drawn from the same population shown in the upper right corner of each sub-plot). 

A more robust quantitative result can be obtained for the full sample. For this we matched the two populations by choosing for every RL quasar 5 RQ quasars with the closest match in $M_{\rm{BH}}$, $L_{5100}$ and redshift. We compare their equivalent width distributions in the left panel of Fig.~\ref{fig:hist} and list some basic statistics on these distributions and their comparison in Table~\ref{tab:stats}. For the full sample we can reject the null hypothesis that both samples are drawn from the same population with high significance,  $p_{\rm{Null}}<10^{-23}$, based on a Kolmogorov-Smirnov test and an Anderson-Darling test. We find a mean offset in the \ewoiii distribution between RL and RQ quasars of $0.164\pm0.034$~dex at 95\% confidence (median of 0.188), i.e. a factor of 1.5. While this difference is not large, it is highly significant, thanks to the large sample from SDSS.

One possible concern on the robustness of this result is on the criteria to define a RL quasar. While our adopted definition is the most widely used for quasars \citep[e.g.][]{Kellermann:1989} it might be too low to select exclusively true RL objects with strong relativistic jets \citep[e.g.][]{Sulentic:2003,Zamfir:2008}. We therefore tested several higher thresholds in $R$, finding consistent results. In particular we here discuss an alternative choice at $R>(R_{\rm{crit}}=80)$, which results in 418 objects classified as RL and therefore the other 7370 as RQ. We carried out the same analysis for this RL criteria and show their \ewoiii results in the middle panels of Fig.~\ref{fig:mmdotplane} and Fig.~\ref{fig:hist} and in Table~\ref{tab:stats}. This criteria further enhances the difference in their \ewoiii distributions, with the RL sample significantly skewed towards high \ewoiii values. We find a difference between RL and RQ by a factor of $1.6-1.8$ ($0.216\pm0.042$, median of 0.258) for this definition of a RL quasar.

\begin{deluxetable*}{lc ccc  ccc  ccc}
\tabletypesize{\scriptsize}
\tablecaption{Statistics of \ewoiii distributions of RL and matched RQ samples}
\tablewidth{18cm}
\tablehead{  &  & \multicolumn{3}{c}{RL} & \multicolumn{3}{c}{RQ} & \multicolumn{3}{c}{RL vs. RQ} \\ \noalign{\smallskip}
\colhead{Sample} & \colhead{N$_{\rm{RL}}$} & \colhead{mean} & \colhead{median} & \colhead{$\sigma$} & \colhead{mean} & \colhead{median} & \colhead{$\sigma$} & \colhead{mean} & \colhead{median}  & \colhead{$p_{\rm{KS}}$}
}
\startdata
\noalign{\smallskip}
$R_{\rm{crit}}=10$ & 746 &  1.34 &  1.36 &  0.43 &  1.18 &  1.17 &  0.37 & $0.164\pm0.034$ & 0.188 & 1.46e-24 \\
$R_{\rm{crit}}=80$ & 418 &  1.40 &  1.44 &  0.40 &  1.19 &  1.18 &  0.38 & $0.216\pm0.042$ & 0.258 & 5.10e-23 \\
$R_{\rm{crit}}(M_\bullet,\dot{M}_\bullet)$ & 394 &  1.41 &  1.45 &  0.40 &  1.23 &  1.21 &  0.39 & $0.187\pm0.044$ & 0.243 & 6.69e-19 \\
$R_{\rm{crit}}=10$ - Pop.B & 525 &  1.36 &  1.38 &  0.42 &  1.21 &  1.19 &  0.36 & $0.147\pm0.039$ & 0.183 & 6.04e-17 \\
$R_{\rm{crit}}=80$ - Pop.B & 287 &  1.44 &  1.45 &  0.38 &  1.22 &  1.21 &  0.38 & $0.218\pm0.048$ & 0.245 & 6.06e-18 \\
$R_{\rm{crit}}(M_\bullet,\dot{M}_\bullet)$ - Pop.B & 324 &  1.42 &  1.45 &  0.39 &  1.24 &  1.23 &  0.39 & $0.182\pm0.047$ & 0.223 & 9.62e-15 \\
$R_{\rm{crit}}=10$ - $R_{\rm{Fe}}$ match & 746 &  1.34 &  1.36 &  0.43 &  1.24 &  1.23 &  0.38 & $0.099\pm0.035$ & 0.124 & 4.91e-13 \\
\enddata
\tablecomments{Statistics for our three applied definitions of a RL quasar for our full sample, for the sample restricted to Population B sources (Pop.B) only and for the additional match in $R_{\rm{Fe}}$. N$_{\rm{RL}}$ gives the number of objects classified as RL in each case. The number of RQ is five times that due to our matching. We list the mean, median and dispersion of the full distribution for both the RL and matched RQ sample. Under RL vs. RQ we also list the difference in their mean, together with its statistical uncertainty, and in the median. The column $p_{\rm{KS}}$ gives the probability that the RL and matched RQ sample are drawn from the same population, based on a Kolmogorov-Smirnov test.}
\label{tab:stats}
\end{deluxetable*}

The radio loudness parameter, defined as a ratio of radio to UV/optical flux, is a phenomenological measure. A more physical parameterization would instead relate this to the intrinsic ratio of jet and accretion power. Both $P_{jet}$ and $P_{acc}$ should scale as $\dot{M}$, so the ratio should be invariant with respect to mass and mass accretion rate, depending only on the ratio of jet and accretion flow efficiencies $\eta_{jet}/\eta_{acc}$.
However, monochromatic radio flux is a non-linear tracer of the jet power \citep{Heinz:2003}. Without any assumption on the boundary conditions from the flow, but with the simple Ansatz that $P_{jet}\propto P_{acc}$ then $f_{6cm}\propto (\mbh\dot{m})^{17/12}$ where $\dot{m}\propto \dot{M}/M_{\rm{BH}}$  \citep{Heinz:2003}, thus $f_{6cm}\propto \dot{M}^{17/12}$. The assumption on the scaling of $P_{jet}$ with accretion power is indeed consistent with observations of dramatically beamed Flat Spectrum Radio Quasars \citep{Ghisellini:2014} as well as of the general RL quasar population \citep{Inoue:2017}.
Furthermore, monochromatic disc power scales with $f_\mathrm{2500}\propto (\mbh\dot{M})^{2/3}$ \citep[e.g.][]{Collin:2002,Davis:2011}. Combining these dependencies, this gives 
\begin{equation}
R_{\rm{crit}}(M_{\rm{BH}},\dot{M})= f_{6cm}/f_{2500}\propto \frac{\dot{M}^{17/12}} {(M_{\rm{BH}}\dot{M})^{2/3}} \propto M_{\rm{BH}}^{-2/3} \dot{M}^{3/4}
\label{eq:rfp}
\end{equation}

We use this new, more physically motivated definition of radio loudness to re-evaluate the trend in \ewoiii shown for the standard definition $R_{\rm{crit}}=10$. For this we define a critical threshold $R_{\rm{crit}}$ for each source above which we classify the quasar as RL which scales like Equation~\ref{eq:rfp}. As reference point for this scaling we set this threshold at $R_{\rm{crit}}(M_{\rm{BH}},\dot{M})=10$ for $\log M_{\rm{BH}}=9.5$ and $\log \dot{M}_{\rm{acc}}=-1.0$. For most other SMBH masses and accretion rates in our sample this then corresponds to a more restrictive value of $R_{\rm{crit}}$, so our sample can still be uniquely classified based on this criteria, where a quasar is RL if $R$ is above $R_{\rm{crit}}$ (given its $M_{\rm{BH}}$ and $\dot{M}$) and RQ if below. This means that at low \mbh and high $\dot{M}_{\rm{acc}}$ the threshold in the optical-to-radio flux ratio is higher and consequently less objects will be classified there as RL compared to the standard radio-loudness criteria, while the number will be approximately the same at high \mbh and low $\dot{M}_{\rm{acc}}$.
This definition reduces the number of RL objects to 395 (and therefore gives 7393 RQ) but also further increases the mean distinction between RL and RQ quasars, as shown in the right panels of Fig.~\ref{fig:mmdotplane} and Fig.~\ref{fig:hist} and listed in Table~\ref{tab:stats}.  We find a difference between the \ewoiii of RL and RQ of a factor of $1.6-1.7$ ($0.187\pm0.044$, median of 0.243) using this alternative, more physical definition of radio-loudness.

We therefore conclude that our results are robust against the detailed definition of radio loudness and choosing a more restrictive or more  physically motivated definition of radio loudness only enhances the discussed trends.

The majority of RL quasars span a more restricted parameter range than RQ quasars in the optical parameter space of the so-called Eigenvector~1 \citep[][see also the discussion below]{Boroson:1992,Sulentic:2000}. In particular they tend to belong to the Population B, defined by \citet{Sulentic:2000} as having FWHM(H$\beta$)$>4000$~km s$^{-1}$ \citep{Sulentic:2003,Zamfir:2008}. Our match in \mbh and $L_{5100}$ also ensures a match in FWHM(H$\beta$) via Equation~\ref{eq:mbhHb}, and therefore our RL and matched RQ sample have the same distribution in respect to FWHM(H$\beta$). Nevertheless we test if our results could be mainly driven by RL quasars which do not belong to Population B. We therefore repeat our analysis restricted to Population B sources only and provide the results in Table~\ref{tab:stats}. They are consistent with those for the full RL sample, verifying that our conclusions are also robust against the restriction to this special regime of the optical parameter space.

\begin{figure}
\centering
\resizebox{\hsize}{!}{\includegraphics[clip]{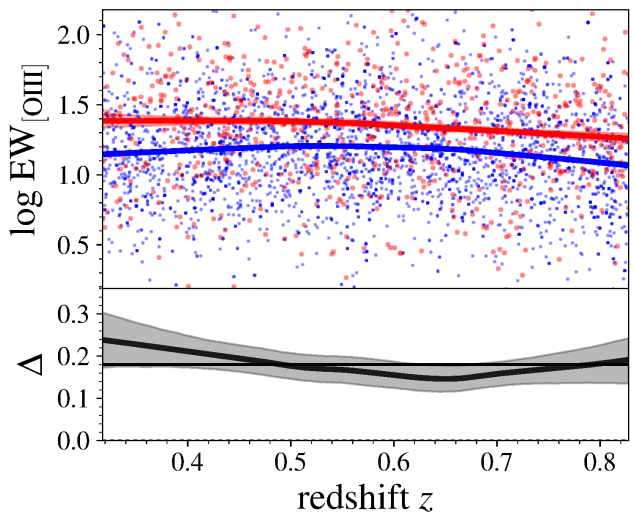}}
\caption{Difference in \ewoiii between the RL (red) and RQ (blue) sample as a function of redshift. Upper panel: The solid line shows the nonparametric local regression result for \ewoiii as a function of redshift for the RL (red) and RQ (blue) sample, while the filled area indicates the 68\% confidence range. The red (RL) and blue (RQ) points show the individual sources in our sub-samples. Lower panel: We show the redshift dependence of the difference in \ewoiii between RL and RQ (gray solid line) and its 68\% confidence interval (gray filled area). The black solid line indicates the mean difference over the full redshift range.
}
\label{fig:ewo3z}
\end{figure}

In Fig.~\ref{fig:ewo3z} we investigate if there is any clear trend in the \ewoiii difference with redshift, were we use our default definition of $R>10$ as radio loudness criteria. 
We use a nonparametric local regression via the LOWESS method to obtain the redshift dependence of \ewoiii for the RL and RQ sample respectively, where we obtain confidence intervals via bootstrapping. Both samples show an indication for mild redshift evolution over the redshift range $0.3<z<0.84$. If true, for our flux limited sample this observed redshift evolution could be a consequence of intrinsic evolution and luminosity effects like the [\ion{O}{3}] Baldwin effect. We here do not attempt to disentangle such potential effects.
However more relevant here is that, as shown in the lower panel of Fig.~\ref{fig:ewo3z}, the difference in \ewoiii is fully consistent with no redshift evolution over the given redshift range.

We did not correct the continuum luminosity $L_{5100}$ for possible host galaxy contribution. However, for the majority of the sample with $\log\,L_{5100}>44.5$ [erg s$^{-1}$] the host galaxy contamination is on average low \citep[$<20$\%;][]{Shen:2011}. It may only affect $L_{5100}$ in the lowest luminosity bins in Fig.~\ref{fig:mlplane} and Fig.~\ref{fig:hist_bins}, where the number of RL sources is low. Even in these bins to first order the host galaxy contribution for RL and RQ should be similar, since they are also matched in \mbh and therefore on average also in $M_*$ via the $\mbh-M_{\rm{bulge}}$ relation \citep[e.g.][]{Kormendy:2013}. Based on the stacked spectra for our RL and RQ sample (see below) we also see no pronounced difference in the Ca H+K absorption feature from the host galaxy component between the RL and matched RQ sample. We conclude that host galaxy contamination should not have any significant influence on our results.

\begin{figure*}
\centering
\includegraphics[width=16cm,clip]{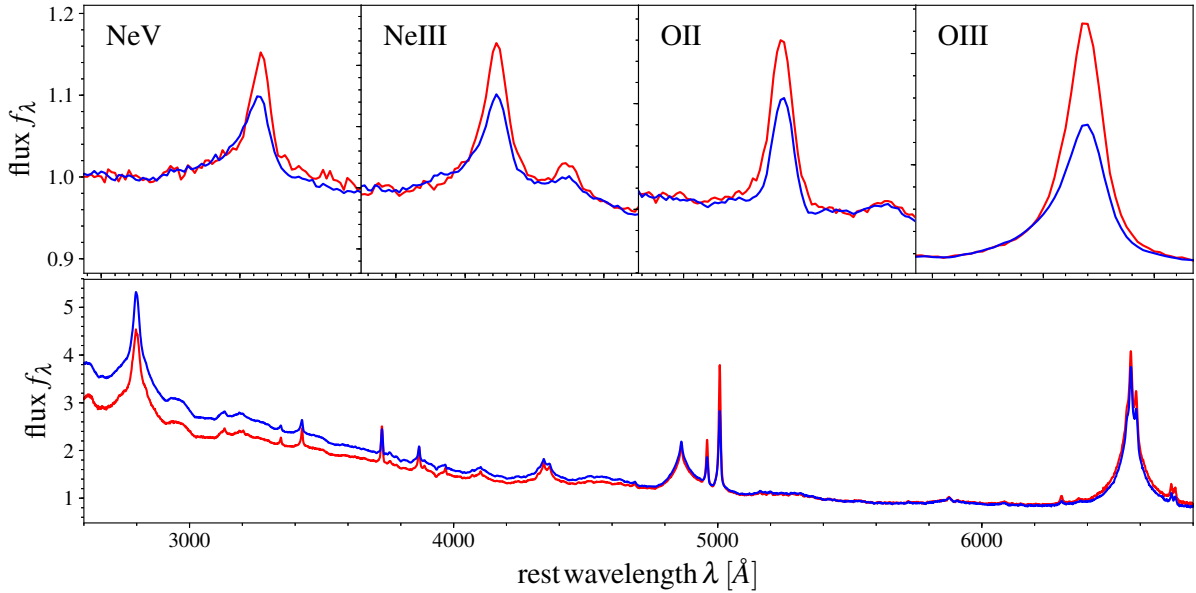}
\caption{
{\bf Lower panel:} Composite spectra of the RL  (red) and matched RQ (blue) quasar sample, where the composite spectra are normalized at 5100\AA{}.
{\bf Upper panel:} Zoom into the composite spectra on the narrow, high ionization lines \ion{Ne}{5}, \ion{Ne}{3}, \ion{O}{2} and \ion{O}{3}. 
}
\label{fig:stackspec}
\end{figure*}

\subsection{RL and RQ composite spectra}
\begin{figure*}
\centering
\includegraphics[width=17.5cm,clip]{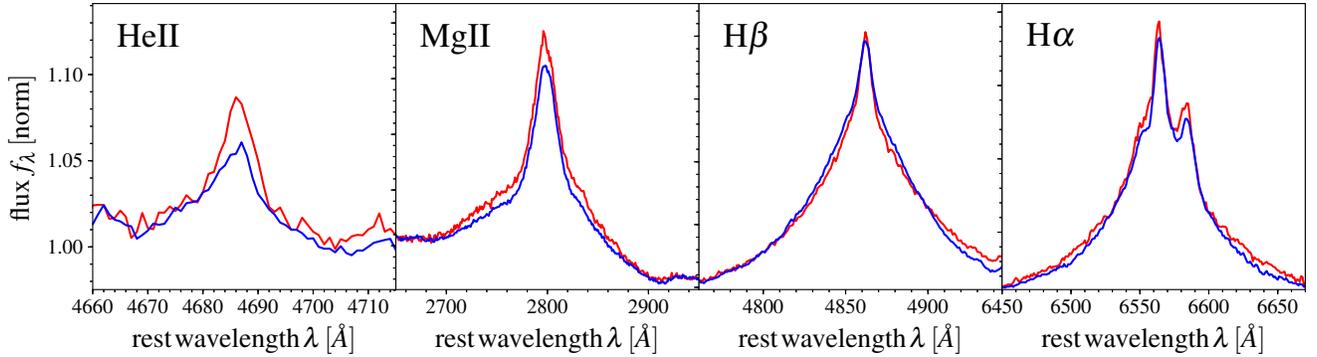}
\caption{Composite spectra of RL  (red) and matched RQ (blue) quasar sample for the broad emission lines \ion{He}{2}, \ion{Mg}{2}, H$\beta$ and H$\alpha$.}
\label{fig:bl}
\end{figure*}

We use the RL and matched RQ sample to generate composite spectra for both matched populations. Each spectrum is shifted into rest-frame \citep[using the redshifts from][]{Hewett:2010}, re-binned to a common wavelength scale and normalized at 5100\AA{} and a stacked spectrum is generated using the median and the geometric mean. In Fig.~\ref{fig:stackspec} we show the full spectrum as well as the high-ionization narrow lines \ion{Ne}{5}, \ion{Ne}{3}, \ion{O}{2} and \ion{O}{3} of the composite for the RL and matched RQ sample, while in Fig.~\ref{fig:bl} we show the broad lower ionization lines \ion{Mg}{2}, H$\beta$ and H$\alpha$ as well as the broad high ionization line of \ion{He}{2}~$\lambda4685$. For \ion{He}{2} and H$\beta$ we have subtracted the optical \ion{Fe}{2} emission line complex in both composites contaminating in particular \ion{He}{2}, using the iron-template for I~Zwicky~1 by \citet{Boroson:1992}.

Focusing on the continuum shape, we find that the RL sample has a redder spectral slope, consistent with an additional reddening of E(B-V)$\sim0.07$ (based on the geometric mean composite). This result is consistent with previous work that found RL quasars to be redder \citep{Brotherton:2001,Labita:2008,Shankar:2016}. We confirm this trend holds also when the samples are being matched in $M_{\rm{BH}}$, $L_{5100}$ and $z$. 
We find a higher Balmer decrement, i.e. the flux ratio between H$\alpha$ and H$\beta$, both for the broad lines and for the narrow lines in the RL composite compared to the RQ composite, supporting the interpretation of larger dust reddening for the RL quasars. We note that for the narrow lines the uncertainties are significant, due to the difficulty to robustly decompose the broad and narrow Balmer lines. We therefore do not apply any dust correction to the narrow line luminosities but emphasize that such a correction will only increase the trend of enhanced [\ion{O}{3}] emission.
Further support for dust reddening is also provided by stronger absorption of the NaID feature present in the RL stack, indicating a larger amount of dust in the host galaxy \citep{Baron:2016}.

\citet{Richards:2003} have shown that dust reddened SDSS quasars show an excess in [\ion{O}{3}] and [\ion{O}{2}] emission compared to the average (low dust content) quasar population. If this excess is inherently related to the fact that these objects are dust reddened and not due to other differences between the sub-samples, like SMBH mass, accretion rate, radio loudness or SMBH spin, this could indicate that the fact that the RL sample is redder contributes to the observed enhancement in [\ion{O}{3}].

We discuss the emission line properties in more detail in Sec.~\ref{sec:disc}.

\section{Discussion} \label{sec:disc}
We have investigated  well-defined samples of RL and RQ quasars, matched in black hole mass and luminosity (or accretion rate). Our main observational result is an on average enhanced [\ion{O}{3}] emission line strength in the RL quasar population at fixed optical continuum luminosity at 5100\AA{}.

We interpret this observed enhancement as being caused by an on average higher EUV luminosity at 35~eV for the RL quasar population, i.e. RL quasars have on average a different SED with a harder ionizing continuum as RQ quasars of the same \mbh and $L_{5100}$.
This then implies a higher bolometric luminosity at fixed mass accretion rate. With Equation~\ref{eq:lbol}  this then directly implies a higher average radiative efficiency and therefore higher SMBH spin. This suggests that the RL quasar population has on average higher black hole spin than the RQ population, in agreement with the prediction from the black hole spin paradigm.

\subsection{Relation to Eigenvector~1}
Enhanced [\ion{O}{3}] emission in RL quasars has been noted before \citep{Marziani:2003}, but we here robustly confirm this trend for a large, well-defined sample controlled for any dependencies on \mbh, luminosity and  Eddington ratio. The trend is most consistently studied in the context of the so-called Eigenvector~1 \citep{Boroson:1992,Sulentic:2000b,Boroson:2002,Shen:2014}. This provides an empirical approach to unify the diversity of quasar spectra in a few principal components, the most important of them being Eigenvector~1. A fundamental property within this Eigenvector~1 is an anti-correlation between the strength of [\ion{O}{3}] and the broad optical iron emission. RL quasars do not span the same range within this Eigenvector space as RQ quasars, but prefer the regime of strong [\ion{O}{3}] and weak iron emission \citep{Boroson:1992,Zamfir:2008}. While Eigenvector~1 itself is a phenomenological quantity, its physical driver has been argued to be the Eddington ratio \citep{Boroson:2002,Shen:2014}. In our study we have fixed  $L_{5100}$ and $M_{\rm{BH}}$ between RL and RQ quasars. Assuming a standard constant bolometric correction for all objects, this would correspond to also fixing the Eddington ratio. Under this assumption we would therefore to first order have controlled for Eigenvector~1. Our matched RL and RQ samples however still show Eigenvector~1 correlations. 
An interesting possibility is that SMBH spin also contributes to Eigenvector~1. 

\begin{figure*}
\centering
\includegraphics[height=6.5cm,clip]{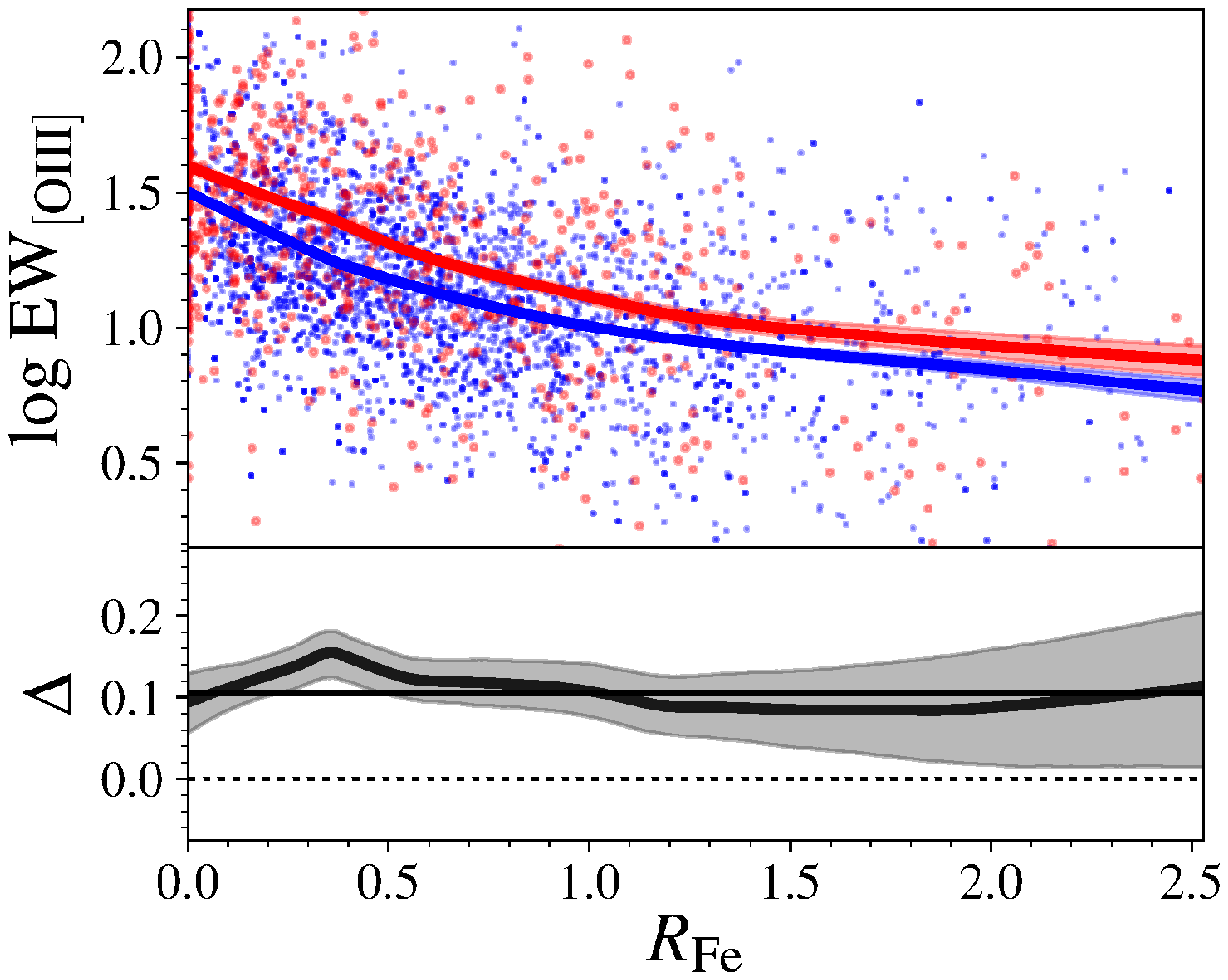} \hspace{0.5cm}
\includegraphics[height=6.5cm,clip]{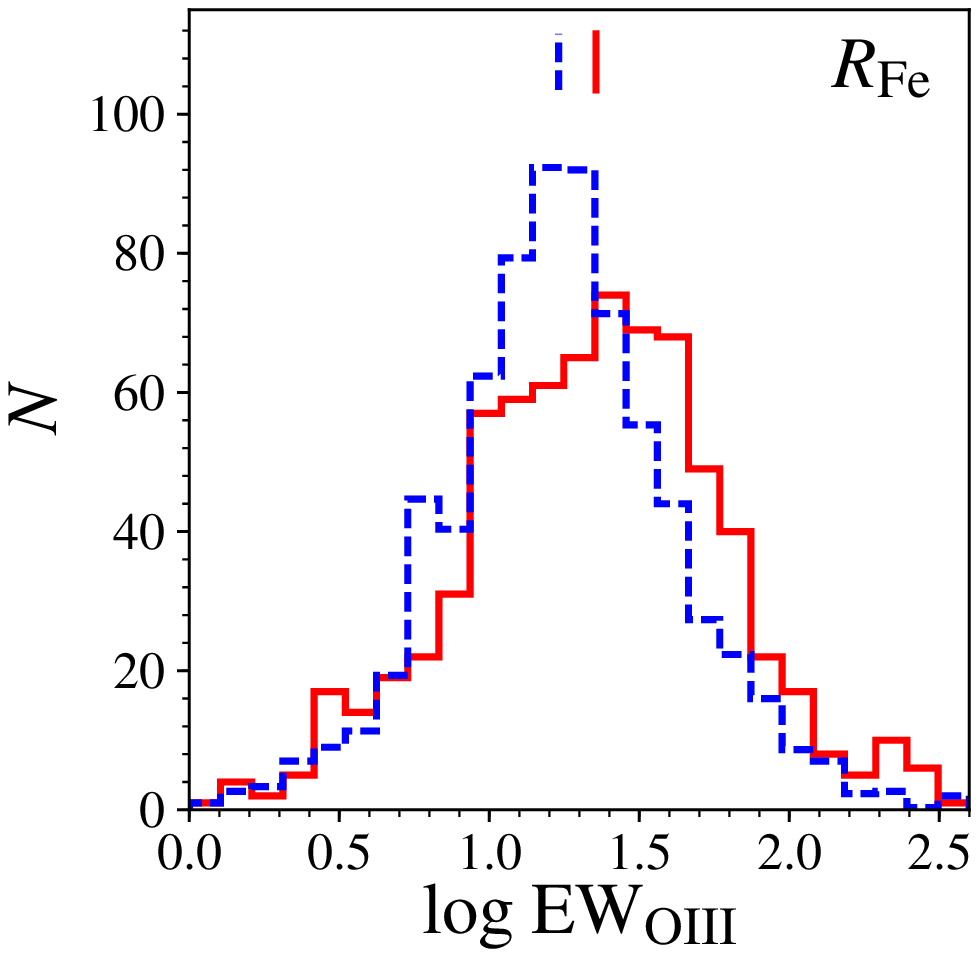} 
\caption{Left panel: We show the correlation between \ewoiii and $R_{\rm{Fe}}$ for the RL (red) and matched RQ (blue) sample, where the points show individual objects, the solid line shows a nonparametric local regression using the LOWESS method and the filled area their 1$\sigma$ confidence interval based on bootstrapping. In the lower panel we show the difference between the RL and RQ sample regression as gray solid line and gray area, where the mean difference over $R_{\rm{Fe}}$ is indicated by the black solid line. Right panel: Histogram of \ewoiii, same as in Fig.~\ref{fig:hist}, but for a RQ sample matched in addition also in $R_{\rm{Fe}}$.}
\label{fig:Rfe}
\end{figure*}

The main observational parameters commonly used to span the Eigenvector~1 parameter space are the FWHM of broad H$\beta$ and the iron strength, given by $R_{\rm{Fe}}=\rm{EW}_{\rm{Fe}}/\rm{EW}_{\rm{H}\beta}$. We have already implicitly matched by FWHM(H$\beta$), but do not control for $R_{\rm{Fe}}$. The main motivation to do so is that we prefer to control for the main physical properties, \mbh and $\dot{M}_\mathrm{acc}$ (assuming our estimates thereof are reasonably accurate), rather than observables whose physical interpretation is not straightforward to understand, like the case for $R_{\rm{Fe}}$. We here explore the consequences of controling for $R_{\rm{Fe}}$, by matching the RL and RQ samples not only in \mbh and $\dot{M}_\mathrm{acc}$, but also in $R_{\rm{Fe}}$  \citep[where $R_{\rm{Fe}}$ is based on the measurements of][]{Shen:2011}. The idea is to compare the RL sample with a sub-population of RQ quasars which occupy the same optical parameter space in Eigenvector~1 as RL quasars, similar to the approach carried out at high-$z$ in \citet{Richards:2011}. The results are shown in Fig.~\ref{fig:Rfe} and Table~\ref{tab:stats}. In the left panel of Fig.~\ref{fig:Rfe} we investigate the dependence of \ewoiii on $R_{\rm{Fe}}$ for both the RL and our matched (in \mbh and $\dot{M}_\mathrm{acc}$) RQ sample with a nonparametric local regression of \ewoiii on $R_{\rm{Fe}}$. While there is a correlation between these parameters for both RL and RQ, their averages are offset by $\sim0.1$~dex (independent of the value of $R_{\rm{Fe}}$). In the right panel of Fig.~\ref{fig:Rfe} we show the \ewoiii distribution of the RL and a RQ sample matched in addition also in $R_{\rm{Fe}}$. As expected, the enhancement in [\ion{O}{3}] is reduced, but maybe surprisingly does not  disappear completely. We find a remaining difference of at least $\sim0.1$~dex, consistent with our results from the left panel of Fig.~\ref{fig:Rfe}. 

We conclude that while we prefer to match in the physical properties \mbh and $\dot{M}_\mathrm{acc}$ only, we note that matching in addition also in $R_{\rm{Fe}}$ reduces the enhancement in [\ion{O}{3}], but still leaves a statistically significant difference of $\sim0.1$~dex.

\subsection{Alternative explanations for the higher  [\ion{O}{3}] EW}
While our interpretation for the enhanced [\ion{O}{3}] emission in RL quasars would provide rare observational support for the black hole spin paradigm, we also have to consider alternative explanations.  Other factors for an enhanced [\ion{O}{3}] luminosity besides differences in the SED and thus ionizing continuum are (1) a different structure (density and/or covering factor) of the NLR, (2) enhanced star formation or (3) line enhancement due to shocks. 

We can directly test some of these scenarios from the composite spectra for our RL quasar and the matched RQ quasar sample. 
In the upper panel of Fig.~\ref{fig:stackspec} we see enhanced emission for RL quasars in all high-ionization narrow lines, with ionization potentials of 97.1~eV, 40.9~eV, 35.0~eV and 13.6~eV for \ion{Ne}{5}, \ion{Ne}{3}, [\ion{O}{3}] and \ion{O}{2} respectively. These lines are all ionized by radiation in the far- to extreme-UV regime. On the other hand, we see no significant enhancement in the low-ionization lines like the Balmer lines as shown in Fig.~\ref{fig:bl}, consistent with the latter responding to lower energy radiation where the RL and RQ samples have the same luminosity. The broad \ion{Mg}{2} line shows a weak enhancement, qualitatively consistent with its moderate ionization potential of 7.6eV.
The broad \ion{He}{2}~$\lambda4686$ line, with a relatively high ionization potential of 24.6eV, shows a pronounced enhancement in the RL stack.

Overall, these trends in the line enhancement for different ionization levels are qualitatively consistent with the SED difference scenario and rule out enhanced star formation as a main driver, since stars are not able to produce an as hard radiation field as required for such high-ionization lines as \ion{Ne}{5}. 

The presence of a clear enhancement in the high ionization broad line of \ion{He}{2} furthermore provides strong support that the line enhancement in [\ion{O}{3}] is not dominated by NLR physics or star formation. 

\begin{figure}
\centering
\resizebox{\hsize}{!}{\includegraphics[clip]{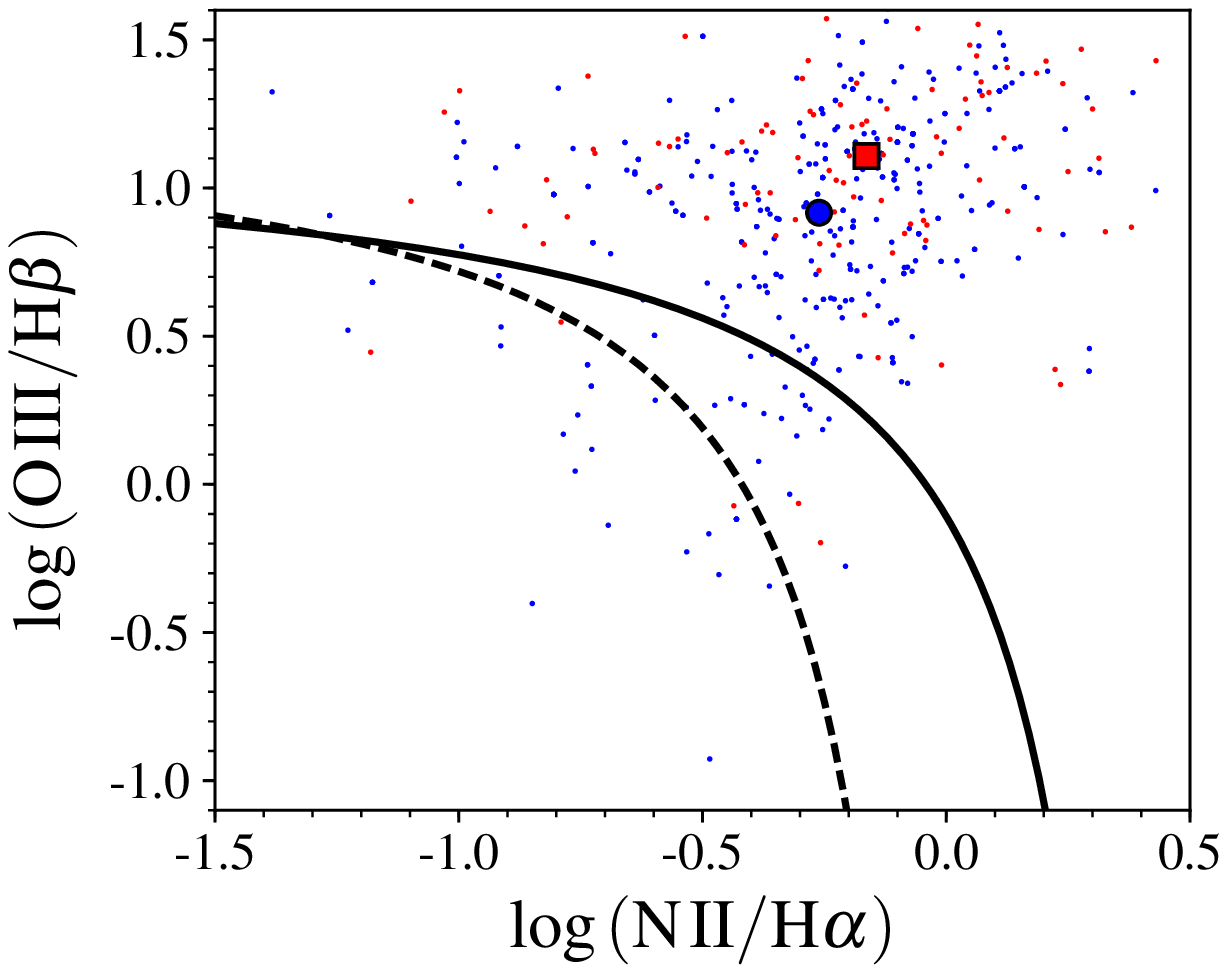}}
\caption{BPT diagnostic diagram for narrow emission lines in the RL (red square) and matched RQ (blue circle) composite spectra and for detections of all four lines in the individual line fits from \citet{Shen:2011} (red (RL) and blue (RQ) dots).
The solid and dashed black lines show the demarcation lines between star forming galaxies and AGN from \citet{Kewley:2001} and \citet{Kauffmann:2003} respectively. Both the RL and RQ composite are located within the region indicating photo-ionization by AGN, while the RL composite is consistent with a harder ionizing spectrum.}
\label{fig:bpt}
\end{figure}

This interpretation is further supported by the BPT diagram  \citep{Baldwin:1981}. This diagnostic emission line diagram is able to distinguish the dominant ionization mechanism of the interstellar medium between AGN and star formation via the narrow emission line ratios of \ion{N}{2}/H$\alpha$ and [\ion{O}{3}]/H$\beta$ \citep{Veilleux:1987,Kewley:2001}. In the high signal-to-noise composite spectra it is possible to disentangle the broad and narrow H$\alpha$ and H$\beta$ components by fitting the respective spectral region with a multi-component model. We measure the narrow H$\alpha$, \ion{N}{2}, H$\beta$ and [\ion{O}{3}] lines from this spectral fit and show their position on the BPT diagram in Fig.~\ref{fig:bpt}. We also show line ratios for individual objects from \citet{Shen:2011}, but caution that they can be more uncertain since the narrow lines are often weak and the de-blending from the broad lines in low to moderate signal-to-noise ratio spectra is challenging. Both the RL and RQ quasars have average line ratios that clearly locate them in the AGN regime. The RL composite is slightly offset from the RQ composite in the direction of a harder ionizing radiation. This provides further support for a harder UV spectrum for RL quasars and does not support enhanced star formation as origin  of the enhanced [\ion{O}{3}] emission.

Another possibility is that the  [\ion{O}{3}] emission is significantly enhanced by the presence of the radio-jet itself, mainly due to jet-driven outflows.
For moderate luminosity radio sources ($L_{\rm rad}<10^{24}$~W Hz$^{-1}$) in particular in Seyfert galaxies it is well known that radio emission, possibly associated with a small scale/compact radio jet, is correlated with disturbed NLR kinematics, indicating the presence of an outflow \citep[e.g.][]{Heckman:1981,Veilleux:1991,Whittle:1992,Husemann:2013,Mullaney:2013}. An alternative suggestion for the origin of this radio emission is synchrotron emission from particles accelerated on the shock front of the radiatively driven outflow itself \citep{Zakamska:2014}. These objects do not posses the strong, large scale relativistic jets seen in RL quasars and in fact are classified as RQ. 
For low-$z$ AGN with high radio luminosity ($L_{\rm rad}\gtrsim10^{24}$~W Hz$^{-1}$) no clear evidence for broadened narrow line widths is seen in large, statistical studies \citep{Mullaney:2013,Coziol:2017}. In high-$z$ powerful radio galaxies, hosting luminous obscured quasars, ionized gas outflows traced by the [\ion{O}{3}] line are frequently seen \citep{Nesvadba:2006,Nesvadba:2008,Nesvadba:2017}, in particular for compact small-scale radio sources \citep{Nesvadba:2007,Kim:2013}. 
While the interpretation that the outflows in these sources are driven by the jet is very plausible, signatures of powerful outflows seem to be common in the general luminous RQ quasar population as well, in particular at high-$z$  \citep[e.g.][]{Netzer:2004,Liu:2013,Brusa:2015,Carniani:2015,Bischetti:2017}. Currently the demographics and fundamental requirements to drive a powerful outflow in an AGN are still not well understood.

Fortunately, we can test the incidence of outflow signatures for our RL sample in comparison to the RQ sample directly from the narrow emission lines in the composite spectra shown in  Fig.~\ref{fig:stackspec}. Focusing in particular on the [\ion{O}{3}] line profile we see that the enhancement is fully due to enhanced core emission, while there is no excess emission in the blue wing component for the RL composite spectrum compared to the RQ composite. We argue that we do not see differences in the line widths and kinematics between the RL and RQ stack. This suggests that the trend is not caused by jet driven outflows or other kinematic components triggered by the presence of the relativistic radio-jet.

Even in those Seyfert galaxies where the NLR kinematics are strongly affected by a small scale jet, the line emission has been found to be still dominated by photo ionization by the AGN rather than shock ionization \citep{Whittle:2005}. Detailed studies of line ratios in powerful radio galaxies also support photoionization by the AGN rather than shocks as the dominant excitation mechanism, both at low redshift \citep{Robinson:1987} and at $z>1.7$ \citep{Villar-Martin:1997}. 
We therefore conclude that for our sample we do not see evidence that the presence of a relativistic jet in RL quasars is responsible for the observed enhancement in [\ion{O}{3}] line emission. 

A remaining uncertainty we cannot fully resolve given current observations lies in our plausible but observationally not well tested assumption of similar average NLR structures (e.g. covering factors) between RL and RQ AGN. This is a priori a reasonable assumption. Detailed IFU studies of mainly RQ and a few RL quasars are generally consistent with this assumption \citep{Husemann:2013}, but the sample sizes are currently limited and thus its validity is far from being observationally well established. Systematic differences in the NLR properties between RL and RQ quasars could in principle be able to mimic the trends we found in this paper, in particular since the NLR structure and physical conditions are the dominant factor for the spread in \ewoiii. However, it is not obvious how such differences should arise. Future studies for larger samples of RL quasars and matched/comparable RQ quasars using spatially resolved spectroscopy will be required to gain information on the presence or absence of any such systematics.
At the least our results indicate some profound differences between RL and RQ quasars beyond black hole mass, luminosity and Eddington ratio which need to be better understood.

However, the clear presence of the line enhancement also in the high-ionization broad line \ion{He}{2}, originating in the broad line region (BLR) and therefore being independent from the NLR properties, argues against systematics in the NLR physics as the sole factor for the [\ion{O}{3}] enhancement. An enhancement in both the narrow and the broad emission lines with high ionization potential cannot be easily explained by intrinsic differences in only the NLR or BLR but rather point to a difference in the ionizing spectrum.

We therefore argue that an intrinsic difference in SED between RL and RQ quasars is the most plausible explanation for the observed enhancement in  [\ion{O}{3}] line emission. The most plausible reason for this SED difference at fixed $\dot{M}_{\rm{acc}}$ is a difference in radiative efficiency, hence black hole spin, with RL quasars having on average higher radiative efficiency and black hole spin than matched RQ quasars.

Alternative models could be the thick disk hypothesis \citep{Tchekhovskoy:2010} or magnetic flux threading \citep{Sikora:2013}. However, it is difficult to compare our interpretation with these models, since these do not provide a precise prediction for the expected SEDs and corresponding emission line strengths, unlike the standard disk assumption we are using here. Furthermore, in the magnetic flux threading scenario there is a priori no reason for RL and RQ quasars to be different in terms of their  [\ion{O}{3}] line emission if the only difference between them at a given \mbh and $\dot{M}_{\rm{acc}}$ is the history of magnetic flux pinned onto the SMBH. On the other hand, the black hole spin scenario provides a consistent straightforward explanation for the observed trends.

\subsection{Implications for the spin, bolometric luminosity and black hole mass of radio-loud and radio-quiet quasars}
For our standard definition of radio-loudness $R>10$ we found a difference between the [\ion{O}{3}] equivalent width of RL and RQ quasars of  a factor 1.5. We interpret this difference as being caused by a difference in SED due to a difference in their average radiative efficiency and thus black hole spin.
For realistic AGN SEDs, the ionizing luminosity traces the bolometric luminosity approximately linearly, which implies the same average difference in bolometric luminosity.
Since $\dot{M}_{\rm{acc}}$ is fixed this also implies the same factor of difference between the radiative efficiencies.

We stress that we do not constrain the individual values of radiative efficiency (and thus black hole spin) for the RL and RQ quasar population. Instead we probe the absolute difference between the two. To illustrate the spin values implied by our results we can explore some reasonable values for the dominant RQ quasar population. We note that the quantitative factor bears uncertainties due to the precise definition of a RL quasar, dust reddening etc. However, as discussed above these will tend to increase the difference between the populations, so we use the factor 1.5 as conservative default value.

This moderate offset disfavors an extreme scenario for the SMBH spin distribution between RL and RQ quasars, where RQ quasars would be non-rotating while RL quasars would be close to maximally spinning, which would result in a larger difference between both populations. Our results rather support wide but different spin distributions for both RL and RQ quasars, where RL quasars are more likely to have high spin. 

Assuming a standard average radiative efficiency of 0.1 for RQ quasars ($a = 0.67$), RL quasars would have an efficiency of 0.15 and thus $a = 0.89$, which is high but not yet close to maximum spin. On the other hand, assuming a black hole spin of $a = 0.2$ ($\epsilon=0.065$) for RQ quasars, as suggested by some theoretical models of spin-evolution  \citep{King:2008}, RL quasars would have an average spin value of $a=0.65$ ($\epsilon=0.097$). Other theoretical models for spin-evolution tend to fall in between \citep{Volonteri:2013}.

We further note that our results may have implications for the determination of black hole mass and bolometric luminosity for RL quasars. It implies that adopting a common bolometric correction factor for RQ quasars will on average underestimate the bolometric luminosity for RL quasars by $\sim0.19$~dex. If the fundamental property in the radius-luminosity relation used for virial black hole mass estimates \citep[e.g.][]{Kaspi:2000,Bentz:2009} is bolometric luminosity rather than $L_{5100}$ (which is used in practice to establish it), virial black hole masses for RL quasars will be according to Eq.\ref{eq:mbhHb} underestimated by $0.5\Delta \log L_\mathrm{bol}\sim0.08-0.09$~dex. We have tested if such a potential underestimation could bias our matched sample construction, by increasing the black hole mass estimate for the RL quasar sample by this average factor. We find our conclusions to be robust against this mass increase.

\section{Conclusions} 
The main observational result of this paper is an average higher [\ion{O}{3}] equivalent width in RL quasars compared to RQ quasars matched in redshift, black hole mass and accretion rate, based on a large well-defined statistical quasar sample from SDSS within $0.3<z<0.84$.
We do not see evidence that the observed trend in [\ion{O}{3}] is driven by star formation or jet-driven outflows. A remaining uncertainty we cannot fully resolve given current observations lies in our assumption of similar average NLR structures between RL and RQ quasars. However, we find a similar enhancement in both narrow and broad high ionization lines (in particular \ion{He}{2}~$\lambda4686$)  which suggests that our result is not driven by NLR physics.

We argue that an intrinsic difference in ionizing continuum, thus in SED, between RL and RQ quasars is the most plausible explanation for the observed [\ion{O}{3}] equivalent width enhancement. We interpret this difference as evidence for on average higher radiative efficiency, hence black hole spin in RL quasars.

The moderate offset in \ewoiii (and by implication in radiative efficiency) of 1.5-1.8 can be explained by a difference in spin between RL and RQ quasars. This could be consistent with the strong spin paradigm, where high spin is a necessary and sufficient condition for the production of relativistic jets, if there is a fairly sharp threshold for jet production. There is some theoretical backing for such a threshold at $a=0.8$ \citep{Maraschi:2012}, in which case our data imply a mean spin of 0.67 (giving a mean efficiency of 0.1) for RQ, while RL would have 0.89. 

However, if the iron line determinations of high spin for local RQ AGN are correct \citep{Reynolds:2014} then high spin is necessary but not a sufficient trigger for jet production. Combining this with our results then points to RL quasars having higher average black hole spin than RQ, but with overlapping distributions. The trigger for the relativistic jet would then be BH spin combined with some other factors such as the accretion
history of magnetic flux \citep{Sikora:2007,Sikora:2013} so that a high spin AGN could be either RL or RQ, while low spin AGN are all RQ.

\acknowledgments
We thank the referee for the constructive comments and suggestions.
A.S. is supported by the EACOA fellowship and acknowledges support by JSPS KAKENHI Grant Number 26800098. C.D. acknowledges support under STFC grant ST/L00075X/1 and a  JSPS Invitation Fellowship for research in Japan (long term) L16518. Y.L. is supported by the National Key Program for Science and Technology Research and Development (No. 2016YFA0400704) and the National Natural Science Foundation of China (No. 11373031). Y.I. is supported by the JAXA international top young fellowship and JSPS KAKENHI Grant Number JP16K13813.

Funding for the SDSS and SDSS-II was provided by the Alfred P. Sloan Foundation, the Participating Institutions, the National Science Foundation, the U.S. Department of Energy, the National Aeronautics and Space Administration, the Japanese Monbukagakusho, the Max Planck Society, and the Higher Education Funding Council for England. The SDSS Web site is http://www.sdss.org/.


\end{document}